\documentclass[aps,prl,twocolumn,superscriptaddress]{revtex4-2}
\usepackage{graphicx}
\usepackage{dcolumn}
\usepackage{bm}
\usepackage{color}
\usepackage{lipsum}
\usepackage{lineno}
\usepackage{subfigure}
\usepackage[colorlinks,urlcolor=blue,linkcolor=blue,citecolor=blue]{hyperref}

\usepackage{siunitx}
\begin{document}

\title{Target density effects on charge tansfer of laser-accelerated carbon ions in dense plasma}%

\author{Jieru Ren} \thanks{These authors have contributed equally to this work.}
\affiliation{MOE Key Laboratory for Nonequilibrium Synthesis and Modulation of Condensed Matter, School of Science,
Xi'an Jiaotong University, Xi'an 710049, China}
\author{Bubo Ma} \thanks{These authors have contributed equally to this work.}
\affiliation{MOE Key Laboratory for Nonequilibrium Synthesis and Modulation of Condensed Matter, School of Science,
Xi'an Jiaotong University, Xi'an 710049, China}
\author{Lirong Liu}
\affiliation{MOE Key Laboratory for Nonequilibrium Synthesis and Modulation of Condensed Matter, School of Science,
Xi'an Jiaotong University, Xi'an 710049, China}
\author{Wenqing Wei}
\affiliation{MOE Key Laboratory for Nonequilibrium Synthesis and Modulation of Condensed Matter, School of Science,
Xi'an Jiaotong University, Xi'an 710049, China}
\author{Benzheng Chen}
\affiliation{MOE Key Laboratory for Nonequilibrium Synthesis and Modulation of Condensed Matter, School of Science,
Xi'an Jiaotong University, Xi'an 710049, China}
\author{Shizheng Zhang}
\affiliation{MOE Key Laboratory for Nonequilibrium Synthesis and Modulation of Condensed Matter, School of Science,
Xi'an Jiaotong University, Xi'an 710049, China}
\author{Hao Xu}
\affiliation{MOE Key Laboratory for Nonequilibrium Synthesis and Modulation of Condensed Matter, School of Science,
Xi'an Jiaotong University, Xi'an 710049, China}
\author{Zhongmin Hu}
\affiliation{MOE Key Laboratory for Nonequilibrium Synthesis and Modulation of Condensed Matter, School of Science,
Xi'an Jiaotong University, Xi'an 710049, China}
\author{Fangfang Li}
\affiliation{MOE Key Laboratory for Nonequilibrium Synthesis and Modulation of Condensed Matter, School of Science,
Xi'an Jiaotong University, Xi'an 710049, China}
\author{Xing Wang}
\affiliation{MOE Key Laboratory for Nonequilibrium Synthesis and Modulation of Condensed Matter, School of Science,
Xi'an Jiaotong University, Xi'an 710049, China}
\author{Shuai Yin}
\affiliation{MOE Key Laboratory for Nonequilibrium Synthesis and Modulation of Condensed Matter, School of Science,
Xi'an Jiaotong University, Xi'an 710049, China}
\author{Jianhua Feng}
\affiliation{MOE Key Laboratory for Nonequilibrium Synthesis and Modulation of Condensed Matter, School of Science,
Xi'an Jiaotong University, Xi'an 710049, China}
\author{Xianming Zhou}
\affiliation{MOE Key Laboratory for Nonequilibrium Synthesis and Modulation of Condensed Matter, School of Science,
Xi'an Jiaotong University, Xi'an 710049, China}
\author{Yifang Gao}
\affiliation{MOE Key Laboratory for Nonequilibrium Synthesis and Modulation of Condensed Matter, School of Science,
Xi'an Jiaotong University, Xi'an 710049, China}
\author{Yuan Li}
\affiliation{MOE Key Laboratory for Nonequilibrium Synthesis and Modulation of Condensed Matter, School of Science,
Xi'an Jiaotong University, Xi'an 710049, China}
\author{Xiaohua Shi}
\affiliation{MOE Key Laboratory for Nonequilibrium Synthesis and Modulation of Condensed Matter, School of Science,
Xi'an Jiaotong University, Xi'an 710049, China}
\author{Jianxing Li}
\affiliation{MOE Key Laboratory for Nonequilibrium Synthesis and Modulation of Condensed Matter, School of Science, Xi'an Jiaotong University, Xi'an 710049, China}
\author{Xueguang Ren}
\affiliation{MOE Key Laboratory for Nonequilibrium Synthesis and Modulation of Condensed Matter, School of Science, Xi'an Jiaotong University, Xi'an 710049, China}
\author{Zhongfeng Xu}
\affiliation{MOE Key Laboratory for Nonequilibrium Synthesis and Modulation of Condensed Matter, School of Science, Xi'an Jiaotong University, Xi'an 710049, China}
\author{Zhigang Deng}
\affiliation{Science and Technology on Plasma Physics Laboratory, Laser Fusion Research Center, China Academy of Engineering Physics, Mianyang 621900, China}
\author{Wei Qi}
\affiliation{Science and Technology on Plasma Physics Laboratory, Laser Fusion Research Center, China Academy of Engineering Physics, Mianyang 621900, China}
\author{Shaoyi Wang}
\affiliation{Science and Technology on Plasma Physics Laboratory, Laser Fusion Research Center, China Academy of Engineering Physics, Mianyang 621900, China}
\author{Quanping Fan}
\affiliation{Science and Technology on Plasma Physics Laboratory, Laser Fusion Research Center, China Academy of Engineering Physics, Mianyang 621900, China}
\author{Bo Cui}
\affiliation{Science and Technology on Plasma Physics Laboratory, Laser Fusion Research Center, China Academy of Engineering Physics, Mianyang 621900, China}
\author{Weiwu Wang}
\affiliation{Science and Technology on Plasma Physics Laboratory, Laser Fusion Research Center, China Academy of Engineering Physics, Mianyang 621900, China}
\author{Zongqiang Yuan}
\affiliation{Science and Technology on Plasma Physics Laboratory, Laser Fusion Research Center, China Academy of Engineering Physics, Mianyang 621900, China}
\author{Jian Teng}
\affiliation{Science and Technology on Plasma Physics Laboratory, Laser Fusion Research Center, China Academy of Engineering Physics, Mianyang 621900, China}
\author{Yuchi Wu}
\affiliation{Science and Technology on Plasma Physics Laboratory, Laser Fusion Research Center, China Academy of Engineering Physics, Mianyang 621900, China}
\author{Zhurong Cao}
\affiliation{Science and Technology on Plasma Physics Laboratory, Laser Fusion Research Center, China Academy of Engineering Physics, Mianyang 621900, China}
\author{Zongqing Zhao}
\affiliation{Science and Technology on Plasma Physics Laboratory, Laser Fusion Research Center, China Academy of Engineering Physics, Mianyang 621900, China}
\author{Yuqiu Gu}
\affiliation{Science and Technology on Plasma Physics Laboratory, Laser Fusion Research Center, China Academy of Engineering Physics, Mianyang 621900, China}
\author{Leifeng Cao}
\affiliation{Advanced Materials Testing Technology Research Center, Shenzhen University of Technology, Shenzhen, 518118, China}
\author{Shaoping Zhu}
\affiliation{Science and Technology on Plasma Physics Laboratory, Laser Fusion Research Center, China Academy of Engineering Physics, Mianyang 621900, China}
\affiliation{Institute of Applied Physics and Computational Mathematics, Beijing 100094, China}
\affiliation{Graduate School, China Academy of Engineering Physics, Beijing 100088, China} 
\author{Rui Cheng}
\affiliation{Institute of Modern Physics, Chinese Academy of Sciences, Lanzhou 710049, China}
\author{Yu Lei}
\affiliation{Institute of Modern Physics, Chinese Academy of Sciences, Lanzhou 710049, China}
\author{Zhao Wang}
\affiliation{Institute of Modern Physics, Chinese Academy of Sciences, Lanzhou 710049, China}
\author{Zexian Zhou}
\affiliation{Institute of Modern Physics, Chinese Academy of Sciences, Lanzhou 710049, China}
\author{Guoqing Xiao}
\affiliation{Institute of Modern Physics, Chinese Academy of Sciences, Lanzhou 710049, China}
\author{Hongwei Zhao}
\affiliation{Institute of Modern Physics, Chinese Academy of Sciences, Lanzhou 710049, China}
\author{Dieter H.H. Hoffmann}
\affiliation{MOE Key Laboratory for Nonequilibrium Synthesis and Modulation of Condensed Matter, School of Science, Xi'an Jiaotong University, Xi'an 710049, China}
\author{Weimin Zhou} \email{zhouwm@caep.cn}
\affiliation{Science and Technology on Plasma Physics Laboratory, Laser Fusion Research Center, China Academy of Engineering Physics, Mianyang 621900, China}
\author{Yongtao Zhao} \email{zhaoyongtao@xjtu.edu.cn}
\affiliation{MOE Key Laboratory for Nonequilibrium Synthesis and Modulation of Condensed Matter, School of Science,
Xi'an Jiaotong University, Xi'an 710049, China}
\bibliographystyle{apsrev4-1}

\date{\today}

\begin{abstract}

We report on charge state measurements of laser-accelerated carbon ions in the energy range of several MeV penetrating a dense partially ionized plasma. The plasma was generated by irradiation of a foam target with laser-induced hohlraum radiation in the soft X-ray regime. We used the tri-cellulose acetate (C$_{9}$H$_{16}$O$_{8}$) foam of 2 mg/cm$^{-3}$ density, and $1$-mm interaction length as target material. This kind of plasma is advantageous for high-precision measurements, due to good uniformity and long lifetime compared to the ion pulse length and the interaction duration. The plasma parameters were diagnosed to be T$_{e}$=17 eV and n$_{e}$=4 $\times$ 10$^{20}$ cm$^{-3}$. The average charge states passing through the plasma were observed to be higher than  those predicted by the commonly-used semiempirical formula. Through solving the rate equations, we attribute the enhancement to the target density effects which will increase the ionization rates on one hand and reduce the electron capture rates on the other hand. 
In previsous measurement with partially ionized plasma from gas discharge and z-pinch to laser direct irradiation, no target density effects were ever demonstrated. For the first time, we were able to experimentally prove that target density effects start to play a significant role in plasma near the critical density of Nd-Glass laser radiation. The finding is important for heavy ion beam driven high energy density physics and fast ignitions.

\end{abstract}

\maketitle
Ion beam interaction with matter is a fundamental process involving complex atomic physics processes. It gets even more complex when the interacting ions and atoms are in a plasma environment. Accurately understanding the details of ion-stopping in plasma is a prerequisite to make use of intense ion beams in high energy density physics \cite{wdm2010,wdm2003,fair,hedp,Deutsch2010}. It requires a correct knowledge of ion charge states when the ion traverses a dense plasma environment \cite{Frank2013,Peter1991,Barriga2016,Ortner2015,olga2005}. It is tempting to relate the charge state of an ion to the parameter Z$_{eff}$ that appears in the stopping power theory, 
\begin{equation}
{-\frac{dE}{dx}\propto \frac{Z_{eff}^{2}}{v^{2}} \times L} 
\end{equation}
where the energy loss $\frac{dE}{dx}$ of heavy ions in the high velocity range is determined by the square of the effective nuclear charge Z$_{eff}$, the Coulomb logarithm $L$ and the ion velocity $v$. Historically, Z$_{eff}^{2}$ is defined as the ratio of the heavy ion stopping S$_{(Z>1)}$ to the stopping of protons S$_{(Z=1)}$ at the same velocity $v$ or at the same energy per nucleon (MeV/u) like S$_{(Z>1)}(v)$ =Z$_{eff}^{2}$  S$_{(Z=1)}(v)$.
The energy loss of the projectile ion in a single collision is determined by the shielding of the nuclear charge \cite{Juaristi,olga2005}. The shielding depends on the impact parameter, the ion charge state, and the population of excited states. In a recent benchmark experiment, we were able to demonstrate the importance of excited ion states for the stopping process \cite{Zhao2021}.

\begin{figure*}
\includegraphics[width=1\linewidth]{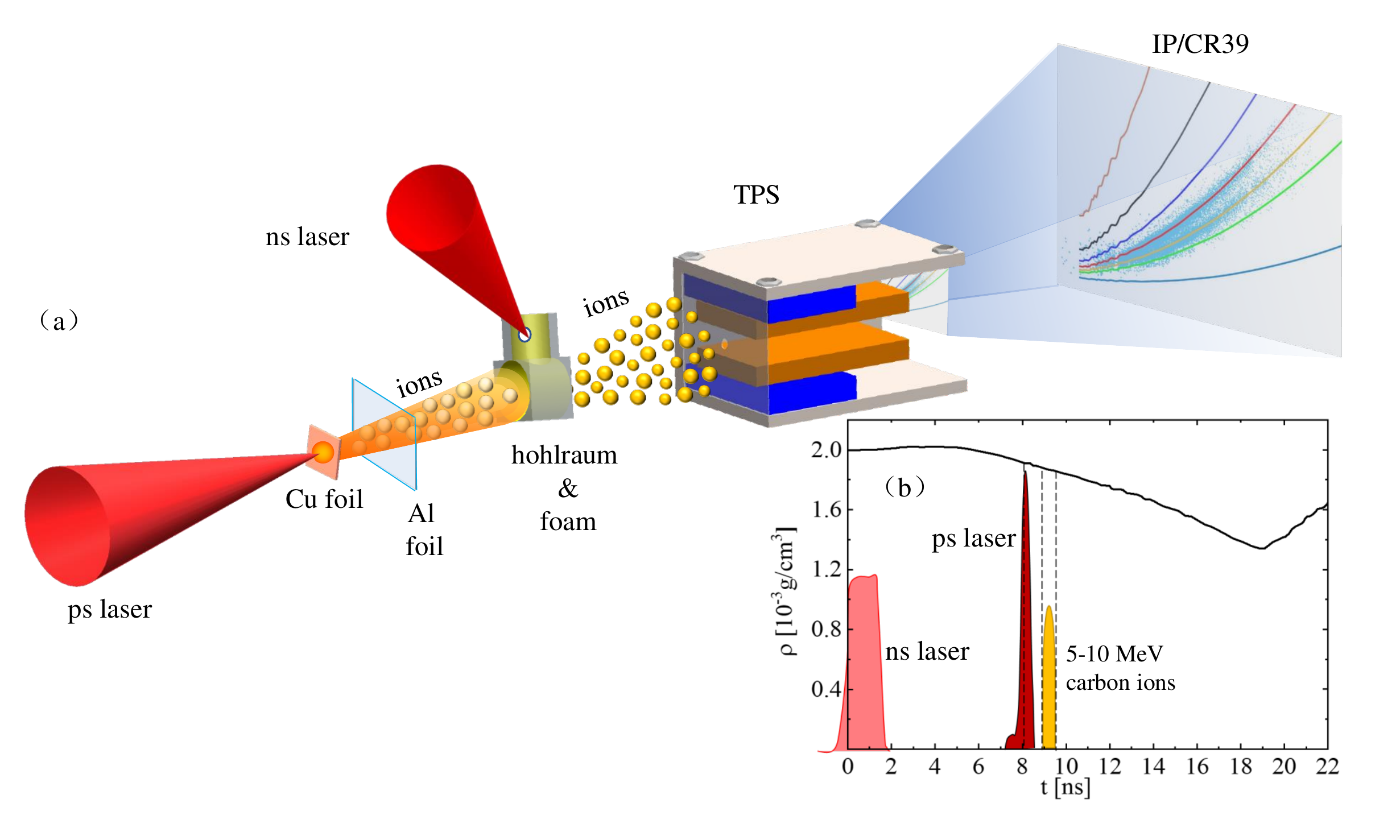}
\caption{\label{fig1} Layout of the experiment. (a) A nanosecond laser is focused onto the inner-wall of the Au hohlraum. The generated X rays from the hohlraum radiation heat the foam, that is attached below the hohlraum, into plasma state. A picosecond laser is focused onto a copper foil, generating an intense short-pulse of carbon ions through the TNSA mechanism. The Al-foil between the copper foil and foam target protects the rear side of the copper foil against backscattering radiation from the ns laser. After a flight-distance of about 1.15 cm, the carbon ions interact with the plasma. The ions passing through the plasma are detected by a Thompson parabola spectrometer coupled to either an image plate or CR39. (b) The time sequence of the ns laser pulse, ps laser pulse and the ion-plasma interaction duration together with the plasma density evolution profile. The data for the density evolution are taken from Ref \cite{Olga2015}.  }
\end{figure*}

When fast heavy ions pass through solid matter, the resulting charge state distribution is shifted to higher charge states compared to the situation when ions pass through the same line density of gaseous material \cite{Lassen1951,Ogawa2007,Olga2007,Olga2004}, the resulting energy loss is higher as well \cite{Geissel1982}. This finding is called density effects and can be well understood within the framework of the Bohr-Lindhard model \cite{BL1954}, where the rapid succession of collisions populates excited states, which on the one hand increases the ionization probability due to the lower binding energy of excited states, and on the other hand decreases the shielding of the nuclear charge. Some experiments with gas-discharge and pinch plasma were performed since 1990s.  Higher charge states are observed when heavy ions penetrate plasma compared to the same amount of cold matter \cite{Dietrich1992,Xuge2017}, because of the lower capture rate of free electrons. This in part also explains the observed enhanced stopping \cite{Jacoby1995,Hoffmann1990,Zhao2021}. In these experiments, the plasma targets were limited in density up to some 10$^{19}$ cm$^{-3}$, and the target density effects were never observed. 

In recent years, the experiments shifted to higher density, where the plasma was created by direct heating of a target foil with an intense laser beam \cite{Cayzac2017,Gauthier2013,Braenzel2018}. However, these plasma targets suffer from the drawback of steep gradients in density and temperature and they are highly transient phenomena. Moreover, traditional optical diagnostic methods are limited to the density corresponding to the critical density of the diagnostic laser. The determination of the properties of the warm dense matter, that is part of the heated target foil, relies strongly on simulations. These factors contribute to  uncertainties in the analysis and the interpretation of the data. In our approach we therefore used a long-living, well-characterized dense plasma. It was created by heating a CHO foam with laser generated hohlraum radiation in the soft x-ray regime \cite{Ma2022, Olga2015, Olga2014, Olga2011}. This plasma sample has been successfully applied in previous experiments to explore the intense proton beam stopping process \cite{Ren2020} and the laboratory observation of white dwarf-like matter \cite{Ma2021}.

Here we employed this plasma sample to measure the charge transfer process of laser-accelerated carbon ions in dense plasma. The plasma life time is approximately 10 ns, which is long compared to the duration of about one picosecond for the pulse length of the laser-accelerated carbon ions. This pulse length in the investigated energy region is longitudinally stretched to about 0.5 ns due to the momentum spread, when it interacts with the target. Therefore, the target conditions can be regarded as constant during the interaction time. This is a significant improvement compared to the situation with direct heated targets and therefore leads to higher precision data. The average charge states of the ions passing through the plasma were compared to theoretical predictions with semiempirical formula and by solving rate equations. These theories significantly misrepresent the experimental data, when target density effects were not included. After modifying the rates of the relevant processes, namely reducing the electron capture rates and increasing the coulomb collision ionization rate, theoretical predictions agree well with our experimental data. 

The experiment was performed at the XG-III laser facility of Laser Fusion Research Center in Mianyang. The experimental layout is displayed in Fig. \ref{fig1}. The ps laser pulse of 130 J energy and 843 fs duration was focused onto a flat copper foil of 10 $\mu$m thickness to generate carbon ions through target normal sheath acceleration (TNSA) mechanism. Protons were accelerated simultaneously, but are not discussed here. To ensure the beam quality from TNSA, a secondary 10 $\mu$m Al foil was inserted to protect the rear side of the copper target from being heated by the nanosecond laser.

The target set-up consists of a gold hohlraum converter (1 mm diameter, 1.9 mm length) with an attached tri-cellulose acetate (C$_{9}$H$_{16}$O$_{8}$) foam (2 mg/cm$^{-3}$ density, 1 mm thickness). The nanosecond laser pulse  of 150 J energy in 2 ns duration was incident upon the inner surface of the hohlraum to generate X rays, which subsequently heated the foam to plasma state. This kind of target scheme and the heating technique allows to generate homogeneous, long-lasting dense plasma, which has been extensively studied at both PHELIX \cite{Olga2015} and XGIII laser facilities \cite{Ma2022, Ma2021, Ren2020}. The current target is the same as the one that was used in the previous experiments and the details of plasma diagnostic can be found there. The plasma temperature was spectroscopically determined to be 17 eV, and the free electron density is 4 $\times$ 10$^{20}$ cm$^{-3}$. 

\begin{figure}
\includegraphics[width=1\linewidth]{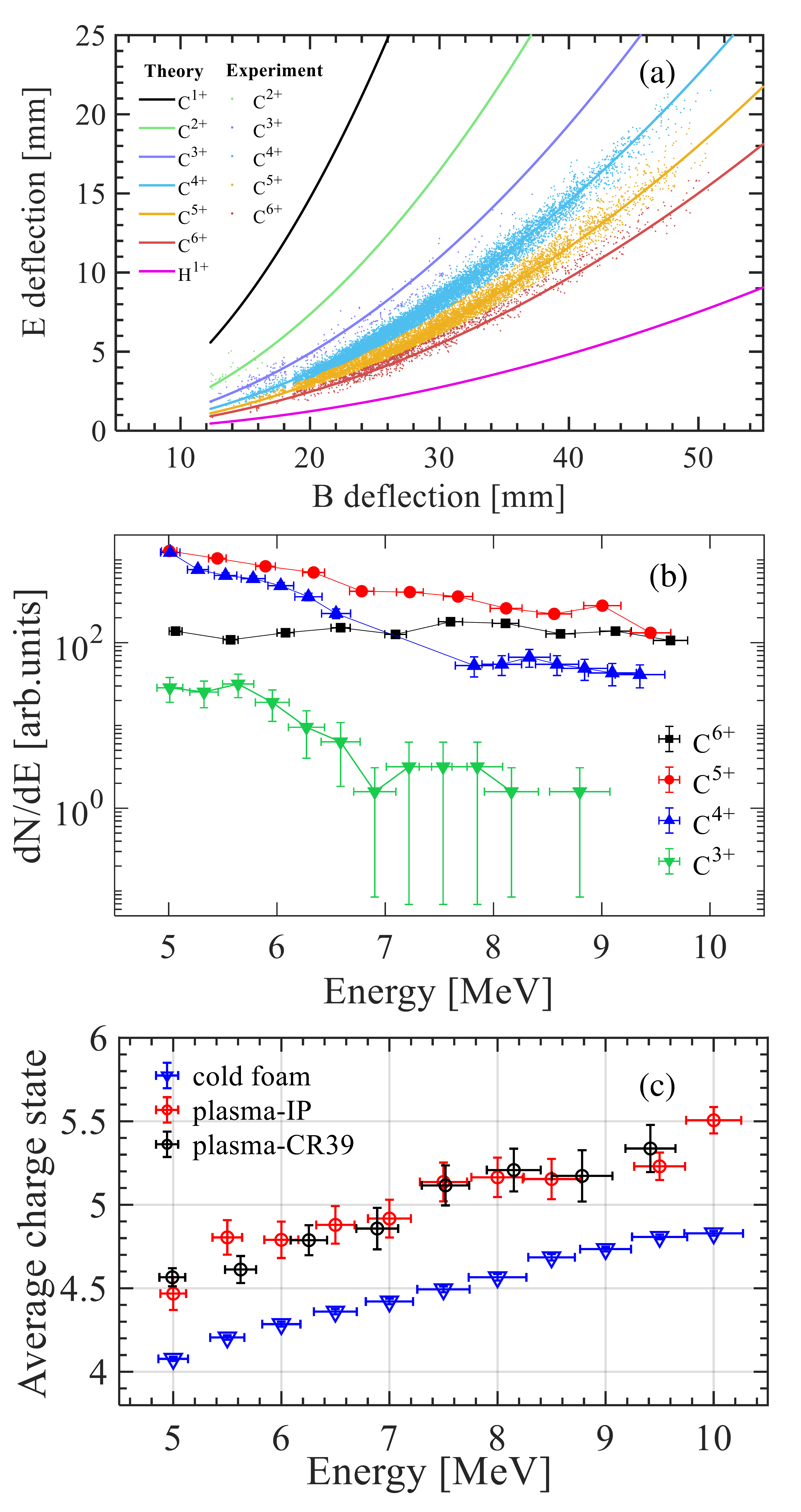}
\caption{\label{fig2} TPS CR39 tracks of carbon ions passing through the plasma and the converted energies spectra as well as the calculated average charge states. (a) TPS CR39 tracks of carbon ions passing through the plasma and the theoretical deflection curves. (b) The converted energy spectra of carbon ions passing through the plasma.  (c) The average charge state for carbon ions penetrating plasma and cold foam. }
\end{figure}
The ps laser was triggered at about 8 ns after the start of the ns laser. The carbon ions were generated instantaneously with the ps laser pulse from the rear side of the copper target. After a flight distance of about 1.15 cm, they reached the foam target. The carbon ions in the investigated range of 5-10 MeV interact with the plasma in the time span between 8.9 ns and 9.4 ns, when the foam was already fully heated, but no macro expansion has occurred yet. The ion-plasma interaction time is on the order of 0.5 ns. This is one order of magnitude shorter than the timescale for the hydrodynamic response of the target. Therefore, the target can be regarded as stationary for our measurement.  

The energies of the carbon ions traversing the plasma were measured with a Thomson Parabola Spectrometer (TPS) coupled to either a Fuji Image Plate or a plastic track detector CR39. The typical tracks of the carbon ions obtained by CR39 are shown in Fig. \ref{fig2}(a) by dots, where the X and Y coordinates represent the magnetic and electric deflection distances relative to the zero order. The solid curves from upper-left to bottom-right are the theoretical deflection distance of ion species from C$^{1+}$ to C$^{6+}$ and H$^{1+}$.  The experimental tracks for protons and the zero order resulting from neutral particles are excluded in this figure. The small gap at the B deflection distance of about 18 mm is due to the unetched border of the CR39.

 The experimental tracks of carbon ions displayed in Fig. \ref{fig2}(a) were converted to energies according to the deflection distance. The energy spectra are shown in Fig. \ref{fig2}(b). The error bars of the intensity in Y direction represent the statistical errors, and the error bars of energy in X direction represent actually the energy resolution of the TPS.
According to the energy spectra in Fig. \ref{fig2}(b), the energy-dependent average charge states
$\overline{Z}(E)$=$\sum_{q}$$P_{q}(E)$$Z_{q}$ are calculated and shown in Fig. \ref{fig2}(c), where $P_{q}(E)$  is the probability of carbon ion in charge state $Z_{q}$ and $E$ is the kinetic energy of carbon. The error bars of the average charge state in Y direction originate from the statistical errors, and the error bars of energy in X direction represent the energy resolution of TPS for ion species that has the lowest resolution.
For comparison, the results of measurements with image plate in another shot are shown as well. The data agree with those measured with CR39 very well. Besides, the average charge states of carbon ions passing through cold foam are also presented. Higher charge states are observed for carbon ions penetrating plasmas compared to that in the cold foam. In this letter, we mainly discuss the results in plasma case.

\begin{figure}
\includegraphics[width=1\linewidth]{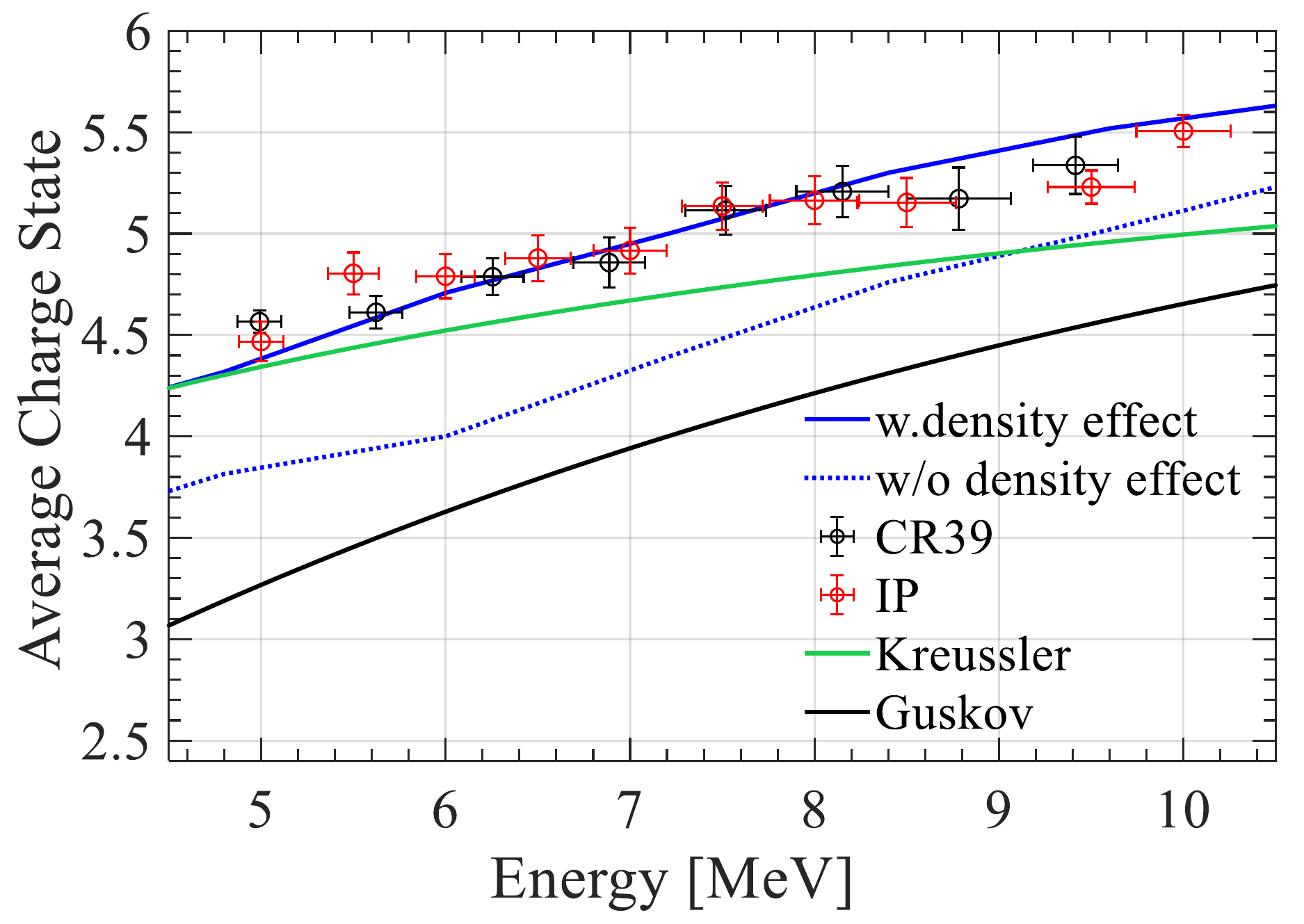}
\caption{\label{fig3} Comparison of the measured average charge state of carbon ions passing through the plasma and the theoretical predictions by analytical formula, solving rate equations with and without target density effects.    }
\end{figure}

According to Morales's model \cite{Morales}, the equilibrium length of the investigated carbon ions in the experimentally-used plasma is about 0.1 mm. This is one order of magnitude lower than the plasma scale. Hence the measured charge states can be reasonably regarded as the  equilibrium charge states of the outgoing carbon ions.
In Fig. \ref{fig3}, the average charge states of carbon ion traversing the plasma were compared to theoretical predictions by the analytical formula proposed by Kreussler \cite{Kreussler} and Guskov \cite{Guskov}. These two models are based on Bohr's stripping criterion \cite{Bohr} while replacing the ion velocity by the average relative ion velocity with respect to the target electrons. The Guskov's model, that describes the plasma electron motion with the Fermi-Dirac function, underestimates our experimental data by about 30\%. The Kreussler's model, that takes into account the thermal velocity distribution of the plasma electrons, predicts higher charge states than Guskov's model, but still underestimates the experimental data. 

In order to understand this discrepancy, we solved the rate equations by taking into consideration all the possible charge transfer processes such as coulomb ionization by plasma ions and free electrons, capture of bound electrons, capture of free electrons, and 3-body recombination processes. The rates for these processes are calculated according to the models reported by Peter and Meyer-ter-Vehn \cite{Peter1991}. As shown in Fig. \ref{fig3}, the calculated equilibrium average charge states without (blue dotted curve) target density effects greatly underestimate the experimental results. Good agreements are achieved between the rate equation predictions (blue solid curve) and experimental data when the the target density effects are taken into account.

\begin{figure}
\includegraphics[width=1\linewidth]{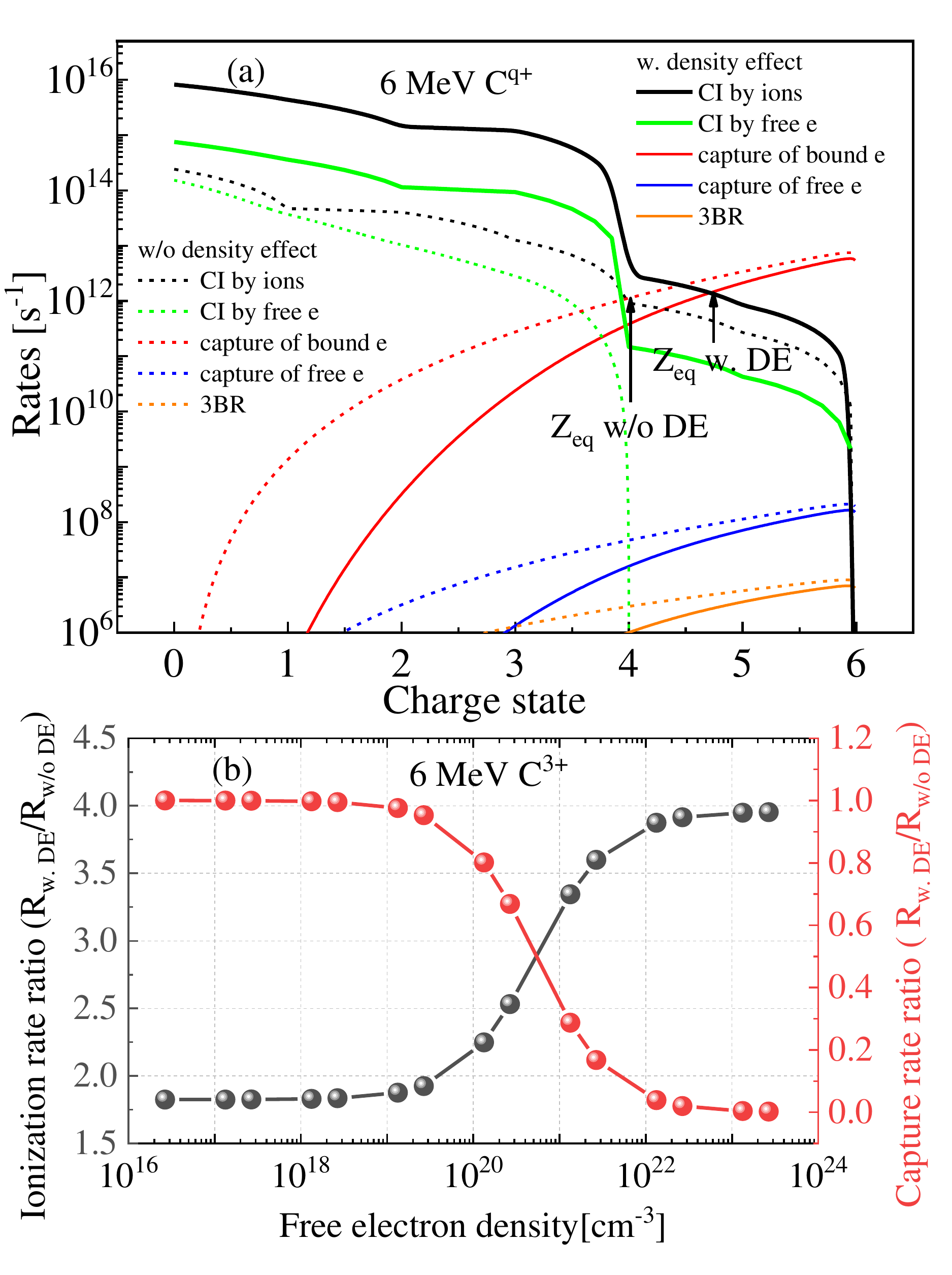}
\caption{\label{fig4} Rate coefficients for charge transfer processes of 6 MeV carbon ions interacting with the experimentally-used plasma. (a) Typical rate coefficients for the Coulomb Ionization (CI) by plasma ions (black curves), Coulomb Ionization by free electrons(green curves), capture of bound electrons(red curves), capture of free electrons or radiative electron capture in another word (blue curves) and three-body recombination (3BR) processes (orange curves) with (solid curves) and without (dashed curves) target density effects; (b) Ratio of the charge transfer rates with and without target density effects modifications versus the target density. }
\end{figure}

The typical rates of the charge transfer processes for 6 MeV carbon ion in the experimentally-used plasma are shown in Fig. \ref{fig4}(a). The dashed curves represent the results in the framework of two-body collisions, where the rate coefficients are density independent. The solid curves represent the results with target density effects modifications. The equilibrium charge state Z$_{eq}$ is achieved when the electron loss rate equals to the electron recombination rate, hence the intersection of the electron ionization and capture curves denotes the value of Z$_{eq}$. The values of the Z$_{eq}$ with and without target density effects are marked in Fig. \ref{fig4}(a). When the target density effects are considered, the average equilibrium charge state is enhanced by about 18\% from 4.0 to 4.7.  The target density effects are discussed in more details below.  

In dense plasmas, the captured electrons, especially those in highly excited states might experience secondary collisions before de-excitation takes place and they therefore can be easily ionized. In this way, the electron capture possibilities are reduced and the ion charge states are enhanced. As shown in Fig. \ref{fig4}(a), when the target density effect is taken into account, considerable reduction occurs for the electron capture processes including bound electron capture, free electron capture and 3-body recombination.

Besides, the frequent succession of collisions in dense plasma gives rise to the two-step processes, excitation and subsequent ionization. This process increases the electron ionization possibilities and leads to the enhancement of the ion charge state. As shown in Fig. \ref{fig4}(a), when the target density effect is considered, the Coulomb ionization rates, either by plasma ions or by free electrons, are drastically enhanced. 

Therefore, we conclude that in the current case of n$_{e}$=4 $\times$ 10$^{20}$ cm$^{-3}$, the target density effects significantly decrease the electron capture rates and increase the electron loss rates. Consequently, the equilibrium average charge states are enhanced. The ratios of the electron capture/loss rates with and without target density effects as a function of free electron density are shown in Fig. \ref{fig4}(b). In the calculations, the plasma was assumed to be partially ionized with the same ionization rates as the experimentally-used sample. The results imply that the target density effects start to play an important role at electron density of about 10$^{20}$ cm$^{-3}$. Our experiments exactly step into this regime, and provide the experimental evidence. 

In summary, the charge states of laser-accelerated carbon ions passing through the dense plasma were measured. By taking advantage of the uniform quasi-static plasma target and short-pulse projectile, high-precision experimental data were obtained. This allows to distinguish between various models. The experimental data exceed the predictions of the Guskov and Kreussler analytical models as well as the rate equation solutions in case that target density effects are not included. The target density effect that will on one hand reduce the electron capture rate and on the other hand increase the coulomb ionization rate, are found to  be responsible for the considerable increase of the charge state in the current case.  Theoretical predictions of rate equations considering both of the two target density effects do well-reproduce our experimental results. To the best of our knowledge, this is the first experiment demonstrating the important role of target density effects at plasma density on magnitude of 10$^{20}$ cm$^{-3}$, which corresponds to about 0.1 percent of solid density. This is important for an accurate understanding of ion-plasma interaction, and is also essential for the physical design of heavy ion beam driven high energy density physics and fast ignitions.

We sincerely thank the staff from Laser Fusion Research Center, Mianyang for the laser system running and target fabrication. The work was supported by National Key R\&D Program of China, No. 2019YFA0404900, Chinese Science Challenge Project, No. TZ2016005, National Natural Science Foundation of China (Grant Nos.12120101005, U2030104, 12175174, and 11975174), China Postdoctoral Science Foundation (Grant no. 2017M623145), State Key Laboratory Foundation of Laser Interaction with Matter (Nos. SKLLIM1807 and SKLLIM2106), and the Fundamental Research Funds for the Central Universities.

\bibliographystyle{unsrt}
\bibliography{ref}

\end{document}